\documentclass[prx,aps,amsfonts,superscriptaddress,amssymb,nofootinbib,floats,twocolumn,10pt]{revtex4-1}
\pdfoutput=1 
\usepackage{graphics}
\usepackage{epsfig}
\usepackage{graphicx}
\usepackage[colorlinks, linkcolor=blue]{hyperref}
\usepackage{amsmath}
\usepackage{amssymb}

\def \ve {\varepsilon}

\def \l {{\vec l}}
\def \ve {\varepsilon}

\def \S {{\cal{S}}}

\def \O {{\cal{O}}}
\def \beq {\begin{eqnarray}}
\def \eeq {\end{eqnarray}}
\def \tn {\textnormal}

\def \el {\ell}
\def \h {\hat}
\def \vp{\varphi}
\renewcommand{\vec}[1]{\boldsymbol{#1}}

\newcommand{\ben}{\begin{equation}}
\newcommand{\een}{\end{equation}}
\newcommand{\ba}{\begin{array}{ccc}}
\newcommand{\ea}{\end{array}}

\begin{document}

\title{Two dimensional spin liquids with $\mathbb{Z}_2$ topological order in an array of quantum wires}

\author{Aavishkar A. Patel}
\email{apatel@physics.harvard.edu}
\affiliation{Department of Physics, Harvard University, Cambridge MA 02138, USA}

\author{Debanjan Chowdhury}
\email{debch@mit.edu}
\affiliation{Department of Physics, Massachusetts Institute of Technology, Cambridge MA 02139, USA}
\date{\today}
\begin{abstract}
Insulating $\mathbb{Z}_2$ spin liquids are a phase of matter with bulk anyonic quasiparticle excitations and ground state degeneracies on manifolds with non-trivial topology. In this paper, we construct a time-reversal symmetric $\mathbb{Z}_2$ spin liquid in two spatial dimensions using an array of quantum wires. We identify the anyons as {\it kinks} in the appropriate Luttinger-liquid description, compute their mutual statistics and construct local operators that transport these quasiparticles. We also present a construction of a fractionalized Fermi-liquid (FL*) by coupling the spin sector of the $\mathbb{Z}_2$ spin-liquid to a Fermi-liquid via a Kondo-like coupling.
\end{abstract}
\maketitle

\section{Introduction}
Mott insulators without any broken symmetries, commonly referred to as quantum spin-liquids (QSL), have been studied theoretically for more than four decades now. Starting with the original theoretical proposal for the resonating valence bond liquid by Anderson \cite{Anderson73}, much of the interest in QSL has been driven by the study of high-temperature superconductivity  \cite{pwa87,GBPWA88,DRSK88,LNW06,DCSSrev} and quantum magnetism in low dimensions \cite{LBrev}. Many interesting insights have been gained by using the notion of topological order \cite{xgwtop}, in order to draw parallels between gapped spin-liquids \cite{rs2,wen1} and other interesting phenomena such as the fractional quantum hall effect \cite{xgw_rev}. 

On the experimental side, a number of quasi-two dimensional materials have been proposed to host QSL ground states \cite{LBrev}. One of the most well studied and promising such materials is Herbertsmithite, consisting of spin-1/2 moments arranged in a kagome lattice. There are indications from theoretical studies that the ground state of the nearest neighbor Heisenberg model (supplemented by next nearest neighbor interactions) on the kagome lattice is a gapped $\mathbb{Z}_2$ spin liquid \cite{SSKagome,swhite11}, even though the question is far from being settled definitively. At the same time, inelastic neutron scattering \cite{YSL12,MPDCSS14} and NMR \cite{Imai15} measurements on Herbertsmithite have detected the existence of a spinon-continuum over a broad energy window and a spin-gap, respectively. 

Theoretical descriptions of QSL ground states usually rely on a `parton' description. Within this prescription, the canonical fermionic operator is fractionalized in terms of excitations that carry its spin and charge separately along with the introduction of an emergent gauge-field that encodes the non-trivial entanglement in the system. The spin-liquid phase corresponds to the deconfined phase of an appropriately defined gauge-theory; examples of gauge groups that often arise in descriptions of various interacting models include e.g. $\mathbb{Z}_2$, $U(1)$, $SU(2)$ coupled to matter fields that are either gapped, gapless at special points, or, gapless along an entire contour in momentum space (see e.g. Refs. \cite{PAL89,TSMPA00,WRXGW,MH04,XGWQO,SSL08} for a few representative examples). There also exist alternative descriptions for time-reversal symmetric QSLs as ground states of exactly solvable (but somewhat artificial) Hamiltonians \cite{Kitaev03,Kitaevhc}, which provide a complementary and useful point of view on the above approaches.

In this paper we take yet another route to arrive at the description of a gapped $\mathbb{Z}_2$ QSL, that does not rely on either of the above two approaches. This approach involves constructing an interacting phase in (2+1)-dimensions starting from a set of decoupled Luttinger liquid wires in (1+1)-dimensions and turning on non-perturbative interactions between the wires. It has been applied remarkably successfully to describe and construct e.g. electronic liquid crystalline phases in doped Mott insulators \cite{KFE98}, the Laughlin state in the fractional quantum hall (FQH) effect \cite{KML02} and more recently even the non-Abelian and compressible FQH states \cite{Teo2014,DM16}. By using this route, we obtain a fully gapped time-reversal symmetric $\mathbb{Z}_2$ QSL and identify the local operators that correspond to and transport the bulk quasiparticles and compute their mutual statistics {\footnote{We note in passing that it is, in principle, possible to construct a gapped spin liquid phase starting from the decoupled ($J_z=0$) and gapless limit of Kitaev's honeycomb model \cite{Kitaevhc} and then studying the effect of a finite $J_z$ perturbatively. We thank A. Vishwanath for pointing this out.}}. A similar approach has been used to construct chiral spin liquids  \cite{KL87} (with broken time-reversal symmetry) \cite{CSL15a,CSL15b}, Abelian topological phases in higher than two spatial dimensions \cite{Meng15, Iadecola16} and even non-Abelian topological spin liquids \cite{CSL15a, Chamon16}.

The rest of this paper is organized as follows: In section \ref{prelim}, we summarize the key features of $\mathbb{Z}_2$ spin-liquids using a Chern-Simons effective field theory description. In section \ref{bos}, we propose a purely bosonic coupled wire construction for the $\mathbb{Z}_2$ spin-liquid, or more specifically the toric-code model, in (2+1)-dimensions. Section \ref{res} summarizes our key results for the bulk quasiparticles, their mutual statistics and the edge physics in the insulating $\mathbb{Z}_2$ spin-liquid within the wire construction. In section \ref{fl*}, we fermionize the above description in order to arrive at a coupled wire construction of a $\mathbb{Z}_2$ fractionalized Fermi liquid (FL*) via a `Kondo'-like construction. We conclude in section \ref{conc} with a summary of our results and an outlook for some of the future directions.

\section{Preliminaries}
\label{prelim}
In this section, we review key features of $\mathbb{Z}_2$ spin-liquids in terms of the low-energy effective theory for its topological states in terms of a Chern-Simons action in imaginary time ($\tau$) \cite{witten89,Freedman04},
\beq
\S_{CS} = \int d\tau~d^2x \bigg[\frac{i}{4\pi} \epsilon_{\mu\nu\lambda}a_\mu^I K_{IJ} \partial_\nu a_\lambda^J + \frac{i}{2\pi} t_I A_\mu  \epsilon_{\mu\nu\lambda} \partial_\nu a_\lambda^I \bigg].\nonumber\\
\eeq
In the above action, $I,J$ are indices extending from $1,..,N$ and $a_\mu^I$ are $N$ U(1) gauge fields, with $A_\mu$ a fixed external `probe' gauge field. The above action realizes an insulating $\mathbb{Z}_2$ spin liquid for $N=2$ with a K-matrix given by
\beq
K=\left( \begin{array}{cc}
0 & 2 \\
2 & 0 \\ \end{array} \right),
\eeq
and where the ground-state degeneracy on a torus is given by $|\tn{det} K|$. The electromagnetic charge of the quasiparticles is determined by the vector $t_I$,
\beq
t_I=\left(\begin{array}{c} 
1 \\ 0 \end{array} 
\right).
\eeq
It is possible to integrate out the internal gauge-fields $\{a_\mu^I\}$, leading to,
\beq
\S_{CS} = (t^T~K^{-1}~t)~\int d\tau~d^2x \bigg[\frac{i}{4\pi} \epsilon_{\mu\nu\lambda} A_\mu \partial_\nu A_\lambda\bigg].
\eeq
The Hall-response is then given by $\sigma_{xy} = (t^T~K^{-1}~t)$ in units of $e^2/2\pi$, which is identically zero for the $\mathbb{Z}_2$ spin-liquid as it preserves time reversal symmetry.

The quasiparticle excitations of the theory are characterized by an integer vector $\l$, such that they couple minimally to the combination: $\sum_I\l_I a_\mu^I$. The self-statistics of a quasiparticle is determined by,
\beq
\theta_\tn{self} = \l K^T \l^{-1},
\eeq
with $\theta_\tn{self}=0(\tn{mod}2\pi)$ for bosons and $\theta_\tn{self}=\pi(\tn{mod}2\pi)$ for fermions. The mutual statistics between two different quasiparticles (`1' and `2') is given by,
\beq
\theta_\tn{mutual} = 2\pi \l_1^T K^{-1} \l_2,
\eeq
with $\theta_\tn{mutual}=\pi(\tn{mod}2\pi)$ for mutual semions. 
In particular, the $\mathbb{Z}_2$ spin-liquid has the following quasiparticle excitations: `$e$', `$m$' and `$\ve$', with
\beq
\l_e=\left(\begin{array}{c} 
1 \\ 0 \end{array}\right),~~ 
\l_m=\left(\begin{array}{c} 
0 \\ 1 \end{array}\right),~~
\l_\ve=\left(\begin{array}{c} 
1 \\ 1 \end{array}\right).
\eeq
It is straightforward to show that $e,~m$ are bosons while $\ve$ is a fermion. All of the above quasiparticles are mutual semions. 

Before we present our construction, the reader might wonder how to obtain a non-chiral and fully gapped state starting from an array of coupled wires. As will become clear in the next section, one of the key ingredients is to be able to find a set of modes with {\it vanishing} self and mutual-commutators. It then allows us to add independent sine-gordon terms for each of these modes in the action, pinning the fields to certain classical values {\it simultaneously} and gapping out the edge excitations~\cite{Lu2012, Teo2014}. 

\section{Bosonic coupled wire construction}
\label{bos}
\begin{figure*}
\begin{center}
\includegraphics[height=2.0in]{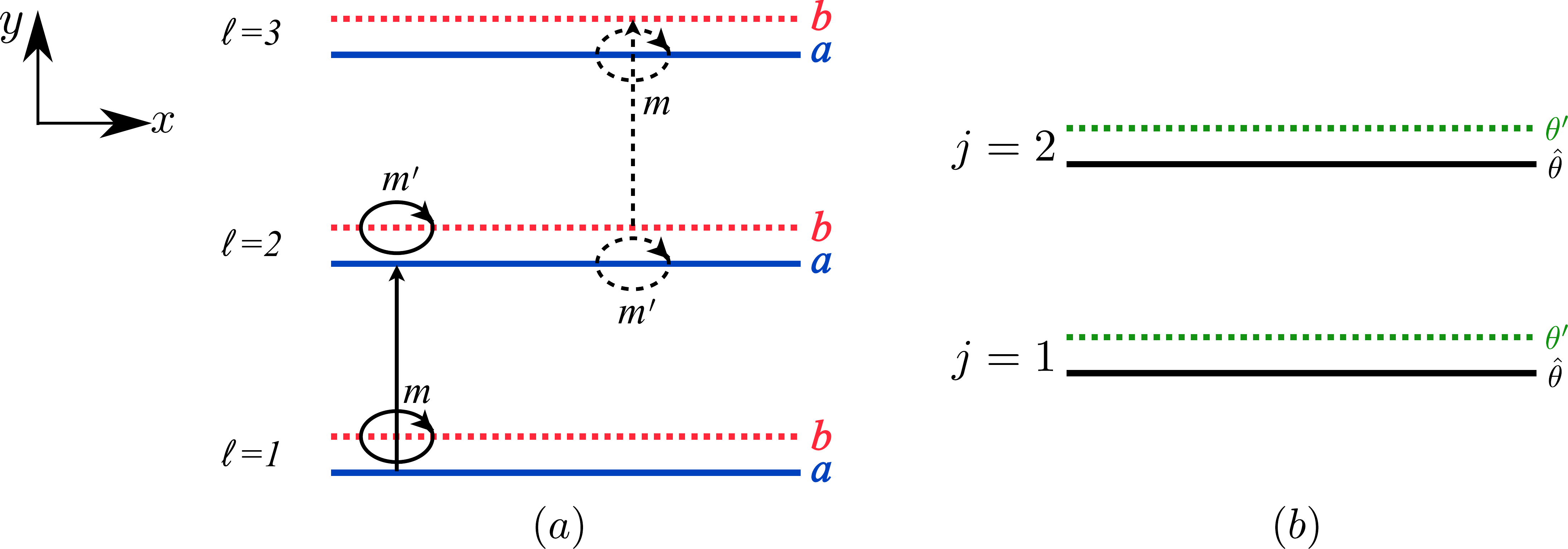}
\end{center}
\caption{(a) Representation of intra and inter-wire scattering terms. Vertical arrows represent the tunneling of bosons between wires (e.g. $\sim \varphi_\el^a-\varphi_{\el+1}^a$). The circular arrows represent backscattering within a wire (e.g. $m\theta_\el^b,~m^\prime\theta_{\el+1}^b$). The dark (dashed) arrows represent a combination of all the processes involved in $\O^\el_1(x)$ ($\O^\el_2(x)$) in Eqn.\ref{ops}.~~(b) Lattice of ``dual" wires labeled by $j$.}
\label{ic}
\end{figure*}

In this section, we arrive at a description of an insulating $\mathbb{Z}_2$ spin liquid using a purely bosonic construction; we defer a discussion of the excitations and edge physics to the next section. We begin by considering an array of uncoupled identical one dimensional quantum wires (labeled: $\el=1, 2, ...$), where each wire consists of two chains: $a$ and $b$ (see Fig.~\ref{ic}a). Each of these chains is described by a {\it nonchiral} Luttinger liquid (LL). We denote the bosonic fields associated with the LL on the $\el$-th wire and on chains `$a$' or `$b$' as: $\{\theta^{a,b}_\el(x),~\vp^{a,b}_\el(x)\}$; they satisfy the following commutation relations,
\beq
[\theta^{a(b)}_\el(x),\vp^{a(b)}_{\el'}(y)] = i\frac{\pi}{2} \tn{sign}(x-y)~\delta_{\el\el'}.
\label{com}
\eeq
We now carry out a series of transformations on the above fields, introducing new degrees of freedom at each stage, as follows. 

We first define a new set of variables,
\beq
\phi_\el^1 &=& \vp_\el^a + m \theta_\el^b,~\phi_\el^2 = \vp_\el^b + m'\theta_\el^a,\nonumber \\
\bar\phi_\el^1 &=& \vp_\el^a - m'\theta_\el^b,~\bar\phi_\el^2 = \vp_\el^b - m \theta_\el^a.
\label{defs1}
\eeq
The commutation relations for the new variables are 
\beq
[\phi_\el^1(x),\phi_{\el'}^2(y)] = -[\bar\phi_\el^1(x),\bar\phi_{\el'}^2(y)] \nonumber\\
= i\frac{\pi}{2} (m+m') ~\tn{sign}(x-y)~\delta_{\el\el'},
\label{com1}
\eeq
which follow trivially from Eqn.~\ref{com}. We also note that the definitions in Eqn.~\ref{defs1} are chosen such that,
\begin{align}
&&[\phi_\el^1(x),\bar\phi_{\el'}^2(y)] = [\bar\phi_\el^1(x),\phi_{\el'}^2(y)]  = 0, \nonumber \\
&&[\phi_\el^1(x),\bar\phi_{\el'}^1(y)] = [\bar\phi_\el^2(x),\phi_{\el'}^2(y)]  = 0. 
\label{com2}
\end{align}
Now introduce a `new' array of wires, defined on the `dual'-lattice sites: $j\equiv\el+\hat{e}_y/2$, with the bosonic fields: $\{\h\theta,\h\phi\}$ and $\{\theta',\phi'\}$ (see Fig.~\ref{ic}(b)). They are defined as,
\beq
\h\theta_j &=& \frac{\phi_\el^1 - \bar\phi_{\el+1}^1}{2},~\h\phi_j = \frac{\phi_\el^2 + \bar\phi_{\el+1}^2}{2}, \nonumber \\
\theta'_j &=& \frac{\phi_\el^2 - \bar\phi_{\el+1}^2}{2},~\phi'_j = \frac{\phi_\el^1 + \bar\phi_{\el+1}^1}{2}.
\eeq
Using the commutation relations in Eqns.~\ref{com1} and \ref{com2}, the commutation relations for the bosonic fields on the dual lattice sites are given by:
\beq
[\h\theta_j(x),\h\phi_{j'}(y)] &=& [\theta'_j(x),\phi'_{j'}(y)] \nonumber\\
 &=& i\frac{\pi}{4} (m+m')~ \tn{sign}(x-y)~\delta_{jj'},
\label{com3}
\eeq
and all other fields commute, i.e.
\beq
[\h\theta_j(x),\theta'_{j'}(y)] &=& [\phi'_j(x),\h\phi_{j'}(y)] \nonumber\\
= [\h\theta_j(x),\phi'_{j'}(y)] &=& [\theta'_j(x),\h\phi_{j'}(y)] = 0.
\label{com4}
\eeq
Thus far we have kept the description in terms of $m,m'(\in\tn{integers})$ completely general. As will become clear later in section \ref{res}, we require from Eqn.~\ref{com3} that $m+m'=4$ in order for the bulk anyonic quasiparticles to be mutual semions with a relative phase of $\pi$. Moreover, in order to make the definitions symmetric, it is natural to choose $m=m'=2$.

The remainder of our discussion will be based on the wires labeled `$j$'. In particular, the usual LL  Hamiltonian for these decoupled wires is given by
\beq
H_0 = \sum_j \frac{\h{v}_j}{2\pi} \int dx \bigg[\frac{1}{\h{g}_j} (\partial_x\h\theta_j)^2 + \h{g}_j (\partial_x\h\phi_j)^2 \bigg] \nonumber\\
+  \sum_j \frac{v'_j}{2\pi} \int dx \bigg[\frac{1}{g'_j} (\partial_x\theta'_j)^2 + g'_j (\partial_x\phi'_j)^2 \bigg],
\eeq
where $\h{v},~v'$ are the effective velocities and $\h{g},~g'$ represent the Luttinger parameters for each individual wire. 

In addition, we also allow for forward-scattering terms between different wires
\beq
H_F = \sum_{j\neq k}\int dx ~(\partial_x\h\phi_j~\partial_x\h\phi_k)~ {\bf \h{M}}_{jk}~ \left(\begin{array}{c} \partial_x\h\phi_j \\ \partial_x\h\phi_k \end{array}\right) \nonumber\\
+ \sum_{j\neq k}\int dx ~(\partial_x\phi'_j~\partial_x\phi'_k)~ {\bf M'}_{jk}~ \left(\begin{array}{c} \partial_x\phi'_j \\ \partial_x\phi'_k \end{array}\right),
\eeq
where the matrices ${\bf \h{M}}_{jk},~{\bf M'}_{jk}$ represent $2\times2$ matrices that describe interactions between wires labelled `$j$' and `$k$'. The theory described in $H_0+H_F$ is quadratic in the fields $\{\h\theta,\h\phi\}$, $\{\theta',\phi'\}$ and describes a `sliding' LL phase. 

Let us now add to the above Hamiltonian further inter-channel scattering terms; it is useful to go back briefly to the description of our system in terms of the original wires labeled `$\el$' (Fig.\ref{ic}a). Then it is possible to write a term of the form
\beq
H_{IC} = \sum_{\el,\alpha=1,2} \int dx ~C_{\el,\alpha}~ \O^\el_\alpha(x), 
\eeq
with two specific choices of $\O^\el_\alpha(x)$:
\beq
\O^\el_1(x) \sim \cos(\phi_\el^1 - \bar\phi_{\el+1}^1) &=& \cos[\vp_\el^a - \vp_{\el+1}^a + m \theta_\el^b + m^\prime \theta^b_{\el+1}] \nonumber\\
&=& \cos(2\h\theta_j), \nonumber \\
\O^\el_2(x) \sim \cos(\phi_\el^2 - \bar\phi_{\el+1}^2) &=& \cos[\vp_\el^b - \vp_{\el+1}^b + m^\prime \theta_\el^a - m \theta^a_{\el+1}] \nonumber\\
&=& \cos(2\theta_j').
\label{ops}
\eeq
The solid and dashed arrows in Fig.~\ref{ic}(a) depict the scattering processes involved above. For our bosonic wires, we need $m, m^\prime \equiv 0~~(\mathrm{mod}~ 2)$ so that the above terms can be written as a combination of inter-wire boson hoppings and scatterings off boson density fluctuations 
\beq
\rho_\ell^{a/b}(x)- \bar{\rho}^{a/b} \sim e^{2i\theta_\ell^{a/b}(x)+2i\pi\bar{\rho}^{a/b} x}
\label{rho}
\eeq 
in the wires~\cite{Teo2014, Lu2012}, where we have taken the average densities $\bar{\rho}^{a/b}$ to be independent of $\ell$. Imagining the wires to be one dimensional lattices with lattice constant $a_0$, we set the average densities of bosons, $\bar{\rho}^{a/b}$, at commensurate values so that the oscillatory factors $e^{i\pi(m\pm m^\prime)\bar{\rho}^{a/b}x}$ are equal to 1. Then, oscillatory factors do not appear in the combinations of hoppings and scatterings used to achieve Eqn.~\ref{ops} and they are thus not trivially rendered irrelevant in the long-wavelength limit.

We note that the scattering terms have been cleverly chosen such that they gap out all possible single-site modes; this follows from the observation that $(a_\ell \varphi_\ell^a + b_\ell \varphi_\ell^b + c_\ell \theta_\ell^a + d_\ell \theta_\ell^b)$ can never commute with all the terms in $H_{IC}$ simultaneously for any non-trivial choice of $a_\el, b_\el, c_\el, d_\el$~\cite{Lu2012}. Hence the bulk of the system will be {\it gapped}---one of the criteria for realizing a $\mathbb{Z}_2$ spin-liquid.

$H_{IC}$ can therefore be most simply expressed as,
\beq
H_{IC} = \sum_j \bigg[ C_{j,1} \cos(2\h\theta_j) + C_{j,2}\cos(2\theta_j')\bigg],
\label{hic}
\eeq
and the entire system is described in terms of the following Hamiltonian,
\beq
H_{SL}[\h\theta,\h\phi,\theta',\phi']=H_0+H_F+H_{IC}.
\label{hsl}
\eeq

By appropriately tuning the values of $\h{g}_j, g'_j$, both the coefficients, $C_{j,\alpha}$, can be made relevant. A simple choice is to set $H_F = 0$ and to set $\hat{g}_j=\hat{g}$ and $g^\prime_j=g^{\prime}$. This choice produces independent sine-Gordon models for each of the `$j$' wires. Then we have the following renormalization group (RG) flow equations for these coefficients ~\cite{ Giamarchi2004, Teo2014}
\beq
\frac{d C_{j,1}}{d l} &=& \left(2 - \frac{m+m^\prime}{2}\hat{g}\right)C_{j, 1},\nonumber \\
\frac{d C_{j,2}}{d l} &=& \left(2 - \frac{m+m^\prime}{2}g^{\prime} \right)C_{j, 2}.
\label{intflow}
\eeq
Therefore at low energies the system flows to a gapped phase in which both $\h\theta$ and $\theta'$ are localized in the respective wells of the cosine potential if $\hat{g}(l=0), g^{\prime}(l=0) < 4/(m+m^\prime)$; this is made possible by the additional fact that these fields commute. We will henceforth make $C_{j,1}$ and $C_{j,2}$ independent of $j$ as well and drop the $j$ label on them. Moreover, in the remainder of this paper, we shall set $m=m'=2$, unless stated otherwise.

\section{Bulk and Edge excitations}
\label{res}
Let us now investigate the nature of the excitations that arise in the system described by $H_{SL}[\h\theta,\h\phi,\theta',\phi']$. In particular, our aim is to identify the bulk anyonic quasiparticles along with the operators that transport them and study the fate of the edge excitations.
\subsection{Bulk quasiparticles and Wilson loops}
It is clear from the form of the term in Eqn.~\ref{hic} that quasiparticles (in the bulk) correspond to kinks in $\h\theta_j$ and $\theta_j'$, where they jump by $\pi$; the states described by $\h\theta_j$ and $\h\theta_j\rightarrow\h\theta_j+\pi$ are energetically equivalent (similarly for $\theta_j'$). The quasiparticle density operators are then given by
\beq
\hat{\rho}_j = \frac{\partial_x\hat{\theta_j}}{\pi},~~\rho^\prime_j = \frac{\partial_x\theta^\prime_j}{\pi}.
\eeq
The operators $e^{\mp i \hat{\phi}_j/2}$ create and annihilate $\hat{\theta}$ quasiparticles while the operators $e^{\mp i \phi^\prime_j/2}$ create and annihilate $\theta^\prime$ quasiparticles respectively. 

Let us now construct the {\it local} operators that hop quasiparticles from wire $j$ to wire $j+1$ 
\begin{align}
&\hat{\Xi}_{j,j+1} = e^{i(\hat{\phi}_j-\hat{\phi}_{j+1}-\theta^\prime_j-\theta^\prime_{j+1})/2} = e^{-2i\theta_{l+1}^a}, \nonumber \\
&\Xi^\prime_{j,j+1} = e^{i(\phi^\prime_j-\phi^\prime_{j+1}-\hat{\theta}_j-\hat{\theta}_{j+1})/2} = e^{-2i\theta_{l+1}^b}, 
\end{align}
which are again proportional to the previously discussed scatterings off density fluctuations on the original $a,b$ wires (Eqn.~\ref{rho}). The operators that transfer quasiparticles from $x_1$ to $x_2$ along wire $j$ are given by
\begin{align}
\hat{\zeta}_j(x_1,x_2) = e^{-i \int_{x_1}^{x_2} dx (\partial_x\hat{\phi}_j)/2}, \nonumber \\
\zeta^\prime_j(x_1,x_2) = e^{-i \int_{x_1}^{x_2} dx (\partial_x\phi^\prime_j)/2},
\end{align}
which can again be expressed in terms of the $a,b$ boson currents and densities.
\begin{figure*}
\begin{center}
\includegraphics[height=2.0in]{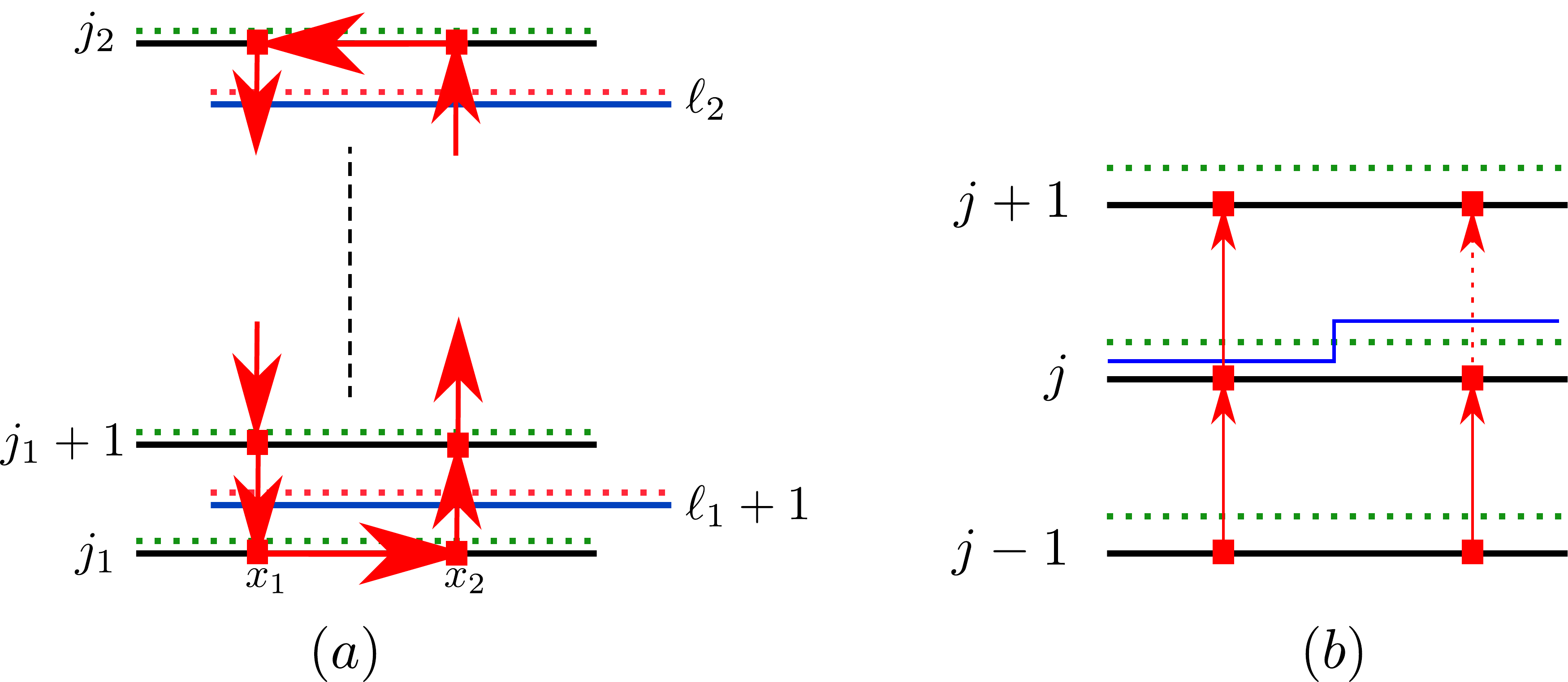}
\end{center}
\caption{(a) Adiabatic transport of quasiparticles gives statistics. Here we take a $\hat{\theta}$ around a loop, picking up a Berry phase proportional to the number of $\theta^\prime$s enclosed.~~(b) The presence of a ``vison" ($\theta^\prime$) induces a branch cut in the hopping of a ``spinon" ($\hat{\theta}$).}
\label{phase}
\end{figure*}
The mutual statistics of the bulk quasiparticles is easily generated by computing the phase generated by taking a quasiparticle around a loop adiabatically~\cite{Teo2014}. Such a process is illustrated in Fig.~\ref{phase}(a). The Berry phases generated by such processes are
\begin{widetext}
\begin{align}
&e^{i\hat{\Phi}} = \left(\prod_{j_1\le j < j_2}\hat{\Xi}_{j,j+1}(x_2)\right)\hat{\zeta}_{j_2}(x_2,x_1)\left(\prod_{j_1\le j < j_2}\hat{\Xi}_{j,j+1}(x_1)\right)^\dagger\hat{\zeta}_{j_1}(x_1,x_2), \nonumber \\
&\hat{\Phi}=\frac{-\int_{x_1}^{x_2}\partial_x\theta^\prime_{j_1}}{2} + \frac{-\int_{x_1}^{x_2}\partial_x\theta^\prime_{j_2}}{2} - \sum_{j_1<j<j_2}\int_{x_1}^{x_2}\partial_x\theta^\prime_{j} = -\frac{\pi}{2}\left(N^\prime_{j_1} + N^\prime_{j_2} + 2\sum_{j_1<j<j_2}N^\prime_{j}\right), \nonumber \\
&\Phi^\prime = -\frac{\pi}{2}\left(\hat{N}_{j_1} + \hat{N}_{j_2} + 2\sum_{j_1<j<j_2}\hat{N}_{j}\right),
\end{align}
\end{widetext}
where $\hat{N}_j$, $N^\prime_j$ are the number of $\hat{\theta}$ and $\theta^\prime$ quasiparticles inside the loop on wire $j$.

A phase of $-\pi$ is picked up for each quasiparticle of the other kind \textit{inside} the loop (and $-\pi/2$ for those on the boundaries of the loop along the wires). Moreover, the phase accumulated is $0$, in the absence of any quasiparticles inside the loop, thereby establishing mutual semionic statistics. We can identify the above quasiparticles as the `$e$' and the `$m$' introduced in section \ref{prelim} above. At this point, there is nothing in our construction that distinguishes between the two quasiparticles.

We note that the above fields commute mutually and hence there can be an additional composite quasiparticle, associated with a \textit{simultaneous} kink in $\h\theta_j,\theta_j'$. The density operator for this quasiparticle is given by $\partial_x(\hat{\theta}_j+\theta^\prime_j)/(2\pi)$ and it is created and annihilated by $e^{\mp i(\hat{\phi}_j+\phi^\prime_j)/2}$ respectively. Repeating the above procedure, we see that this quasiparticle has semionic statistics with each of the $\hat{\theta}$ and $\theta^\prime$ quasiparticles. Similarly, we also note that each such quasiparticle within the loop contributes a phase of $-2\pi$ to the one being taken around. Since this involves a full revolution, it implies an exchange statistical angle of $-\pi$, corresponding to only half a revolution. Thus, this additional composite quasiparticle is a fermion and can be identified as the `$\varepsilon$' introduced earlier in section \ref{prelim}. 

\begin{figure}
\begin{center}
\includegraphics[width=\columnwidth]{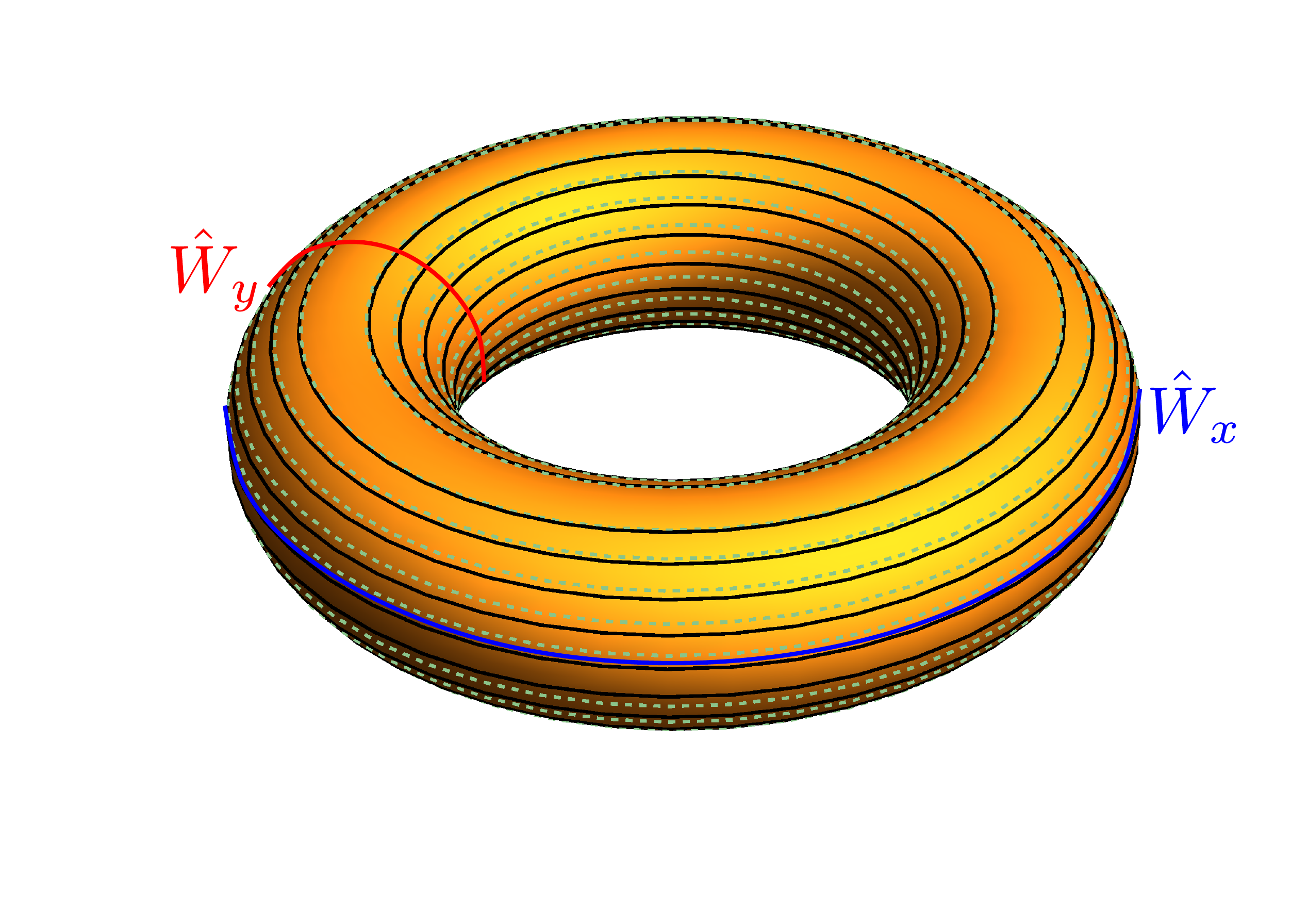}
\end{center}
\caption{Wire array of Fig.~\ref{ic}(b) on a torus. The Wilson loop operators $\hat{W}_y$ and $\hat{W}_x$ are shown in red and blue respectively.}
\label{torus}
\end{figure}

We now place the array of $n$ wires (i.e. $j=1,..,n$) on a torus of dimensions $(L_x, n)$ (Fig.~\ref{torus}). The Wilson loop operators~\cite{witten89,Sagi2015,DCSS16} are then given by,
\begin{align}
&\hat{W}_y(x) = \prod_{j=1}^n \hat{\Xi}_{j,j+1}(x),~~\hat{W}_x = \hat{\zeta}_1(0,L_x), \nonumber \\
&W^\prime_y(x) = \prod_{j=1}^n \Xi^\prime_{j,j+1}(x),~~W^\prime_x = \zeta^\prime_1(0,L_x).
\end{align}
We use periodic boundary conditions to identify $n+1\equiv 1$ and $L_x \equiv 0$. They obey the algebra
\beq
\hat{W}_x W^\prime_y(x) = -W^\prime_y(x)\hat{W}_x,\nonumber\\
W^\prime_x \hat{W}_y(x) = -\hat{W}_y(x)W^\prime_x,
\eeq
with all other combinations commuting. This operator algebra is easily realized by 2 independent sets of Pauli matrices, signaling the 4-fold degeneracy of the ground state on the torus.

One possible choice for the action of the time reversal operator $\mathcal{T}$ on the bosonic fields is~\cite{Lu2012}
\beq
\mathcal{T}: \theta_\ell^a \rightarrow \theta_\ell^a,~\theta_\ell^b\rightarrow -\theta_\ell^b,~\varphi_\ell^a\rightarrow -\varphi_\ell^a,~\varphi_\ell^b\rightarrow \varphi_\ell^b + \pi.\nonumber\\
\label{timerev}
\eeq
The wires labeled `$b$' can then be thought of as being derived from the bosonization of XX spin-$1/2$ chains~\cite{Giamarchi2004,SubirsBook}, while the wires labeled `$a$' are simply neutral spinless bosons. This is consistent with the above choice of time reversal, and the requirement that $e^{4i\pi\bar{\rho}^b x}$ not oscillate. Then, the inter-wire terms in Eqn.~\ref{ops} are only invariant under the $SO(2)$ rotations of the spin components in the XY plane, given by $\varphi^b_\ell\rightarrow\varphi^b_\ell+f$, and not under the $SO(3)$ rotations that mix $\varphi^b_\ell$ and $\theta^b_\ell$ \cite{MBEB}. 

This choice of time reversal sends $\hat{\theta}_j\rightarrow-\hat{\theta}_j$ while leaving $\theta^\prime_j$ invariant; since we wish to identify the time reversal odd excitations of the toric-code with physical spin densities, we call $\hat{\theta}$ ``spinons" and $\theta^\prime$ ``visons".  With this choice, the kink/antikink creation operators transform as 
\begin{align}
&\mathcal{T}:~e^{\pm i\hat{\phi}_j/2}\rightarrow \mp i e^{\mp i \hat{\phi}_j/2},~~e^{\pm i \phi^\prime_j/2}\rightarrow e^{\pm i \phi^\prime_j/2},  \nonumber \\
&\mathcal{T}^2:~e^{\pm i\hat{\phi}_j/2}\rightarrow - e^{\pm i\hat{\phi}_j/2},~~e^{\pm i \phi^\prime_j/2}\rightarrow e^{\pm i \phi^\prime_j/2},
\end{align}
keeping in mind that $\mathcal{T}$ involves complex conjugation. Thus spinon kinks get switched to anti-kinks and vice versa, since the physical spin density is odd under $\mathcal{T}$. It also follows that $\mathcal{T}^2 = -1$ for the `$\varepsilon$' (fermion) quasiparticle.

Since the bulk is gapped we can perturbatively add inter-wire hoppings for the spinons to the hamiltonian,
\begin{align}
&H^{\mathrm{hop}}_{j,j+1}(x)= -t e^{i\Phi_{j,j+1}(x)} e^{i(\hat{\phi}_j-\hat{\phi}_{j+1})/2} + \mathrm{h.c.} \nonumber \\
&\Phi_{j,j+1}(x) = -(\langle\theta^\prime_j(x)\rangle+\langle\theta^\prime_{j+1}(x)\rangle)/2.
\label{hop1}
\end{align}
As long as $t$ is much smaller than the bulk gaps, this should not destabilize the coupled-wire fixed point. Let us consider the hopping of a spinon at $x$ from wire $j-1$ to $j+1$, in the presence of a vison located at $x=0$ on wire $j$ (Fig.~\ref{phase}(b)). The hopping amplitude for this process is $h_{j-1,j+1}(x) \propto t^2 e^{i(\Phi_{j-1,j}(x)+\Phi_{j,j+1}(x))}$. Since the presence of the vison causes $\theta^\prime_j$ to jump by $\pi$ at $x=0$, we can see that $h_{j-1,j+1}(x>0)=-h_{j-1,j+1}(x<0)$. Thus the vison induces a branch cut for the spinon hopping, as we know already from parton constructions of $\mathbb{Z}_2$ spin liquids~\cite{Huh2011}. A different model of coupled spin chains without $\mathbb{Z}_2$ topological order but with spinons capable of hopping between chains was previously proposed in Ref.~\cite{Nersesyan03}.     

\subsection{Bulk gap}

The bulk quasiparticle excitations are gapped, with a finite energy required to create them. The gaps are non-universal and are in general different for $\hat{\theta}$ and $\theta^\prime$, which would correspond to different gaps for the spinons and visons. As we show below, they depend on the details of the renormalization group flows of the sine-Gordon models on the `$j$' wires. The flow equations for $C_1,~C_2$ are given by Eqn.~\ref{intflow}; the equations for $\h{g},~g^\prime$ are ~\cite{ Giamarchi2004}
\beq
\frac{d \hat{g}}{dl} = -  A_1 C_1^2 \hat{g}^3, \nonumber \\
\frac{d g^\prime}{dl} = -  A_2 C_2^2 g^{\prime 3},
\eeq
where $A_1$ and $A_2$ are nonuniversal numerical constants. Defining $z^\parallel_1 = 2\hat{g} - 2$, $z^\parallel_2 = 2g^{\prime} -2$, $z^\perp_1=C_1/\sqrt{8A_1}$ and $z^\perp_2=C_2/\sqrt{8A_2}$, we have the Kosterlitz-Thouless RG equations for small $|z^\parallel_{1,2}|$,
\begin{align}
&\frac{dz^\parallel_{1,2}}{dl}\approx-(z^{\perp}_{1,2})^2, \nonumber \\
&\frac{dz^\perp_{1,2}}{dl}=-z^\parallel_{1,2}z^\perp_{1,2}.
\end{align}
When $z^\parallel_{1,2} < 0$ and $z^\perp_{1,2} \le  |z^\parallel_{1,2}|$, the system flows to strong coupling and the bulk is gapped. Additionally, when $z^\parallel_{1,2}>- 1$, the low energy excitations in the bulk are the kinks we discussed previously, and the bulk gaps $\Delta_{1,2}\sim\sqrt{z^\perp_{1,2}/(z^\parallel_{1,2}+2)}$~\cite{Giamarchi2004}.  Note that the RG equations do not have a stable fixed point, and hence the flows will be stopped by non-universal scales. The kinks are of the form, 
\begin{align}
&\langle\hat{\theta}(x)\rangle\sim \tan^{-1}(x/w_1), \nonumber \\
&\langle\theta^\prime(x)\rangle\sim \tan^{-1}(x/w_2),
\end{align} 
where $w_{1,2}\sim 1/\sqrt{z^\perp_{1,2}(z^\parallel_{1,2}+2)}$~\cite{Giamarchi2004}. However, we will assume that the widths $w_{1,2}$ of the kinks are much smaller than the other length scales in our model, and treat the kinks as sharp step-functions.

\subsection{Physics at the edges}
In a wire array where $\ell$ runs from $1$ to $n$, the fields $\bar{\phi}^1_1,\bar{\phi}^2_1$ and $\phi^1_n,\phi^2_n$ living on the edges do not appear in the sine-Gordon terms $\mathcal{O}^\ell_\alpha$. Thus, these non-chiral modes are gapless and also commute with the Hamiltonian. In the absence of additional symmetries, we are free to add sine-Gordon terms to the edges to gap these modes out; for example, we can add
\beq
H_{\mathrm{edge}} = \int dx\left(D_1 \cos(2\bar{\phi}^1_1) + D_n \cos(2\phi^1_n)\right),
\label{edgeH}
\eeq
and tune the kinetic terms on the edges, as we did before for the bulk, to make these relevant. This localizes $\bar{\phi}^1_1$ and $\phi^1_n$. Due to the non-vanishing commutators (Eqn.~\ref{com1}) between $\bar{\phi}^1_1,\bar{\phi}^2_1$ and $\phi^1_n,\phi^2_n$, fluctuations of $\bar{\phi}^2_1$ and $\phi^2_n$ are maximized due to the uncertainty principle, and these modes are consequently gapped~\cite{Lu2012}. Note that $\bar{\phi}_1,\bar{\phi}_2$ (or $\phi^1_n,\phi^2_n$) cannot be simultaneously localized owing to their non-vanishing mutual commutators (Eqn.~\ref{com1}). Thus, our results are consistent with the usual expectations for the edge of the toric code, which can either be of the $m$ or $e$ type, but not both \cite{kitaevb,Barkeshli14}.

The edge fields $\bar{\phi}^1_1\rightarrow -\bar{\phi}^1_1$ and $\phi^1_n\rightarrow -\phi^1_n$ under time reversal (Eqn.~\ref{timerev}), thus if they are localized to $0$, by Eqn.~\ref{edgeH} the edges will be gapped without spontaneously breaking time reversal symmetry. 

\section{Fermionization and $\mathbb{Z}_2$ Fractionalized Fermi Liquid}
\label{fl*}
The previous section provided a coupled-wire construction for the $\mathbb{Z}_2$ spin-liquid using a purely bosonic model. Let us now fermionize the spinons in $H_{SL}$, as this will be necessary for our construction of the $\mathbb{Z}_2$ FL*. The FL* is a phase of matter where a Fermi-liquid with gapless excitations coexists with a background spin-liquid. The simplest examples of FL* arise in two-band Kondo-Heisenberg lattice models \cite{Senthil2003}. In a simplified picture of such models, the local moments interacting via Heisenberg exchange interactions can form the spin liquid, while the conduction electrons form a Fermi liquid with a `small' Fermi surface. In the limit of a weak Kondo exchange between the local and itinerant electrons, the resulting FL* phase violates Luttinger's theorem \cite{MO00}, which can be understood as arising from the presence of background topological order \cite{APAV04}. 

In order to fermionize the spinons, we first add a new set of bosonic ``chargon" fields $\theta^c_j,\phi^c_j$ to the wires labeled by `$j$' which satisfy 
\beq
[\theta^c_j(x),\phi^c_{j'}(y)] = i\pi~ \tn{sign}(x-y)~\delta_{jj'}.
\label{com5}
\eeq
Their Hamiltonian is given by
\beq
H_c &=& \sum_j \left[\frac{v^c}{2\pi}\left(\frac{1}{g^c}(\partial_x\theta^c_j)^2+g^c(\partial_x\phi^c_j)^2\right) + C_c \cos(2\theta^c_j) \right], \nonumber \\ 
\eeq
with $g_c$ chosen so that $H_c$ is gapped, and $H_{SL}\rightarrow H_{SL}+H_c$. We then consider the $\hat{\theta},\hat{\phi}$ and $\theta^c,\phi^c$ to respectively describe the long-wavelength spin and charge sectors of spinful fermionic Luttinger liquid wires,
\beq
\hat{\theta}_j = \theta_{j\uparrow}-\theta_{j\downarrow},~~\hat{\phi}_j = \phi_{j\uparrow}-\phi_{j\downarrow}, \nonumber \\
\theta^c_j = \theta_{j\uparrow}+\theta_{j\downarrow},~~\phi^c_j = \phi_{j\uparrow}+\phi_{j\downarrow}.
\eeq

Given the commutation relations in Eqns.~\ref{com3}, \ref{com4}, and \ref{com5} we demand that the fields introduced above satisfy the following commutation relations:
\beq
[\theta_{j\sigma}(x),\phi_{j'\sigma'}(y)] = i\frac{\pi}{2}~ \tn{sign}(x-y)~\delta_{jj'} \delta_{\sigma\sigma'}.
\eeq
These are the canonical Luttinger liquid commutators. Thus, the fermion creation and annihilation operators may then be written as ($\sigma=\uparrow,\downarrow$),
\beq
\psi^R_{j\sigma}(x) &=& \frac{F_{j}}{\sqrt{2\pi x_c}}~ e^{i[k_F^0x + \phi_{j\sigma}(x) + \theta_{j\sigma}(x)]},\nonumber \\
\psi^L_{j\sigma}(x) &=& \frac{F_{j}^\dagger}{\sqrt{2\pi x_c}}~ e^{i[-k_F^0x + \phi_{j\sigma}(x) - \theta_{j\sigma}(x)]}.
\label{fermions}
\eeq
The $F_j$ represent the Klein factors that ensure anticommutation on different wires and $x_c$ is a short-distance cutoff; one possible choice for the Klein factors is~\cite{Teo2014}
\begin{align}
&F_j = (-1)^{\sum_{\sigma,l<j}(\mathcal{N}^R_{l\sigma}+\mathcal{N}^L_{l\sigma})}, \nonumber \\
&\mathcal{N}^{R/L}_{j\sigma} = \pm \int \frac{dx}{2\pi}~\partial_x(\phi_{j\sigma}(x)\pm\theta_{j\sigma}(x)).
\end{align}
Under time reversal given by Eqn.~\ref{timerev} we have
\begin{align}
&\theta_{j\uparrow}\leftrightarrow\theta_{j\downarrow} \nonumber \\
&\phi_{j\uparrow}\rightarrow-\phi_{j\downarrow}+\pi/2,~~\phi_{j\downarrow}\rightarrow-\phi_{j\uparrow}-\pi/2,\nonumber \\
&\psi_{j\sigma}^{R/L}\rightarrow (-1)^\sigma i \psi_{j\bar{\sigma}}^{L/R},
\end{align}
where we made a symmetric choice for the phase factors in the second line of the above.

The spin lowering and raising operators corresponding to the above definitions are given by,
\beq
S_j^+ &=& \frac{1}{2}(\psi_{j\uparrow}^{R\dagger} \psi_{j\downarrow}^R + \psi_{j\uparrow}^{L\dagger} \psi_{j\downarrow}^L),\nonumber \\
S_j^- &=& \frac{1}{2}(\psi_{j\downarrow}^{R\dagger} \psi_{j\uparrow}^R + \psi_{j\downarrow}^{L\dagger} \psi_{j\uparrow}^L),
\eeq
which can be re-expressed in terms of the bosonic fields as,
\beq
S_j^+ &=& \frac{1}{4\pi x_c} \bigg[e^{-i(\h\phi_j + \h\theta_j)} + e^{-i(\h\phi_j - \h\theta_j)} \bigg], \nonumber \\
S_j^- &=& \frac{1}{4\pi x_c} \bigg[e^{i(\h\phi_j + \h\theta_j)} + e^{i(\h\phi_j - \h\theta_j)} \bigg].
\eeq
On the other hand, the $z-$component is given by,
\beq
S^z_j = \frac{1}{2}(\psi_{j\uparrow}^{R\dagger} \psi_{j\uparrow}^R - \psi_{j\downarrow}^{R\dagger} \psi_{j\downarrow}^R + \psi_{j\uparrow}^{L\dagger} \psi_{j\uparrow}^L - \psi_{j\downarrow}^{L\dagger} \psi_{j\downarrow}^L)=\frac{\partial_x\hat{\theta}_j}{(2\pi)}. \nonumber\\
\eeq
Thus we have $\mathbf{S}_j\rightarrow-\mathbf{S}_j$ under $\mathcal{T}$. 

$S_\pm$ switch anti-kinks ($\equiv~\downarrow$) to kinks ($\equiv~\uparrow$) and vice versa. Thus, they create and annihilate two spinons at a time respectively. 
The spin sector sine-Gordon term maps to the backscattering term
\beq
C_1\cos(2\hat{\theta}_j) \rightarrow \tilde{C}_1 \psi^{L\dagger}_{j\uparrow}\psi^R_{j\uparrow}\psi^{R\dagger}_{j\downarrow}\psi^L_{j\downarrow} + \mathrm{h.c.}~,
\eeq
which does not have any oscillatory $e^{2ik_F^0x}$ factors and hence is not trivially rendered irrelevant in the long-wavelength limit. For the charge sector we have
\beq
C_c\cos(2\theta^c_j) \rightarrow e^{4ik_F^0x} \tilde{C}_c \psi^{L\dagger}_{j\uparrow}\psi^R_{j\uparrow}\psi^{L\dagger}_{j\downarrow}\psi^R_{j\downarrow} + \mathrm{h.c.}~
\eeq
Imagining the fermions to live on a lattice with lattice constant $a_0$ as before, we tune to half filling $k_F^0 = \pi/(2a_0)$ to eliminate the oscillatory factor in the above so that we can have both charge and spin gaps.

Thus we have,
\begin{widetext}
\begin{align}
&H_{SL}\rightarrow H_f + H_v, \nonumber \\
&H_f =  \sum_{j,\sigma}\int dx~\Bigg[v_f\left(\psi^{R^\dagger}_{j\sigma}\left(-i\frac{\partial}{\partial x}\right)\psi^R_{j\sigma}-\psi^{L^\dagger}_{j\sigma}\left(-i\frac{\partial}{\partial x}\right)\psi^L_{j\sigma}\right) + \tilde{C}_1 (\psi^{L\dagger}_{j\uparrow}\psi^R_{j\uparrow}\psi^{R\dagger}_{j\downarrow}\psi^L_{j\downarrow} + \mathrm{h.c.})+ \tilde{C}_c (\psi^{L\dagger}_{j\uparrow}\psi^R_{j\uparrow}\psi^{L\dagger}_{j\downarrow}\psi^R_{j\downarrow} + \mathrm{h.c.}) \Bigg], \nonumber \\
&H_v =  \sum_j \frac{v'_j}{2\pi} \int dx \bigg[\frac{1}{g'} (\partial_x\theta'_j)^2 + g' (\partial_x\phi'_j)^2 \bigg] + C_2 \cos(2\theta^\prime_j).
\end{align}
\end{widetext}
where $H_v$ corresponds to the vison piece unaffected by the fermionization.

The spinons (together with chargons) may be hopped between wires by adding perturbative nonchiral hoppings of the fermions 
\begin{align}
&H^{\mathrm{hop},f}_{j,j+1}(x) = \nonumber \\
&-t e^{i\Phi_{j,j+1}(x)}(\psi^{L\dagger}_{j+1,\sigma}(x)\psi^{L}_{j\sigma}(x)+\psi^{R\dagger}_{j+1,\sigma}(x)\psi^{R}_{j\sigma}(x))+\mathrm{h.c.},
\label{hop2}
\end{align}
with the phase of the hopping amplitude given by Eqn.~\ref{hop1} as the chargons have trivial mutual statistics with the visons.

\begin{figure*}
\begin{center}
\includegraphics[height=2.0in]{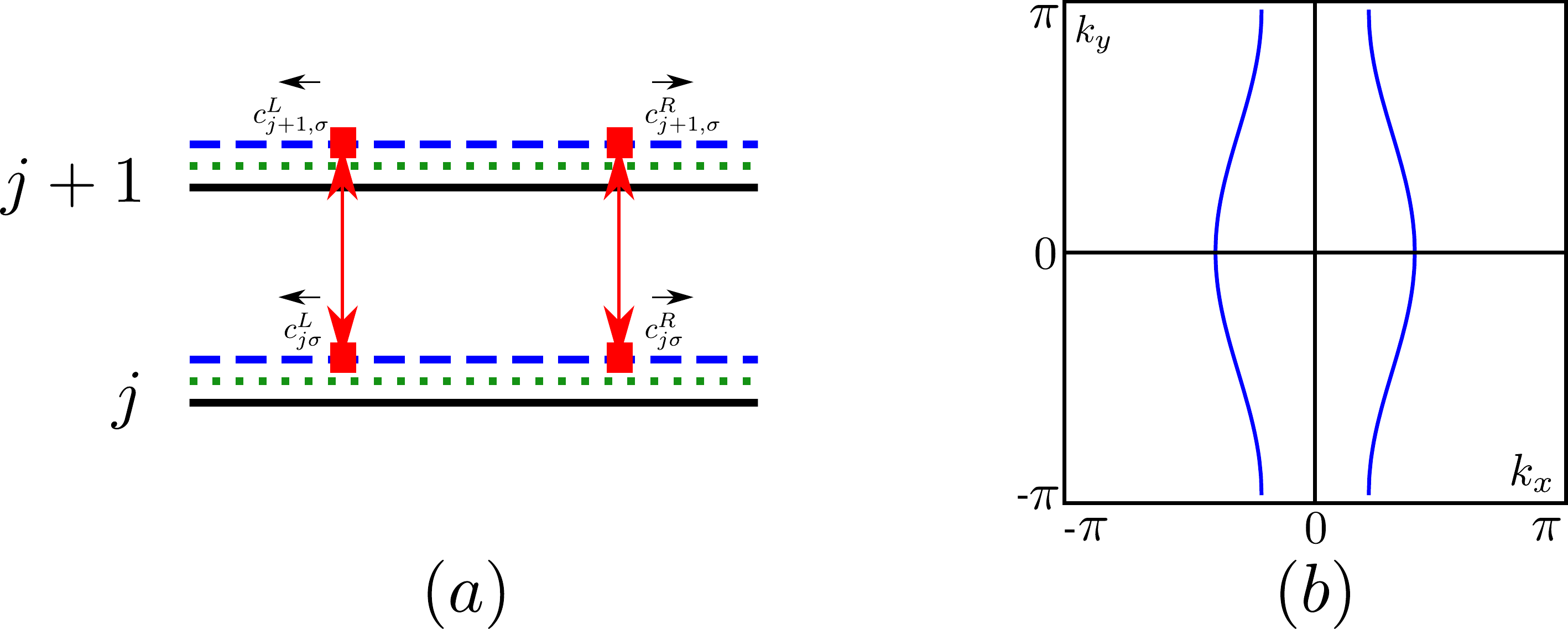}
\end{center}
\caption{(a) The additional set of wires (dashed, blue) carrying the conduction electrons. Non-chiral tunneling between these wires (red arrows) allows the electrons to form a two dimensional Fermi liquid, coupled to the spin liquid background (FL*). (b) Schematic Fermi surface of such a Fermi liquid.}
\label{cwire}
\end{figure*}

To realize the FL*, we add another set of wires to the sites labeled by `$j$', carrying the conduction electrons labeled by $c^{R/L}_{j\sigma}$; we also add non-chiral hoppings between these wires so that the electrons form a quasi-1d Fermi surface (Fig.~\ref{cwire}). The conduction electrons are described by
\begin{align}
&H_{el} = \sum_{j,\sigma}\int dx~\Bigg[v_F\left(c^{R^\dagger}_{j\sigma}\left(-i\frac{\partial}{\partial x}\right)c^R_{j\sigma}-c^{L^\dagger}_{j\sigma}\left(-i\frac{\partial}{\partial x}\right)c^L_{j\sigma}\right) \nonumber \\
&-t_1 (c^{R\dagger}_{j+1,\sigma}c^R_{j\sigma}+c^{L\dagger}_{j+1,\sigma}c^L_{j\sigma}+\mathrm{h.c})\Bigg].
\end{align}
The spin-density corresponding to the conduction electrons is denoted,
\beq
\mathbf{s}_j = \frac{1}{2}(c_{j\sigma}^{R\dagger}\boldsymbol{\tau}_{\sigma\sigma^\prime} c_{j\sigma^\prime}^R + c_{j\sigma}^{L\dagger}\boldsymbol{\tau}_{\sigma\sigma^\prime} c_{j\sigma^\prime}^L).
\eeq
We now couple the spin sector of the electrons to the spinons via a local spin-spin coupling, similar to the Kondo coupling used in the original description of the FL*~\cite{Senthil2003}. We analyze two different cases below {\footnote {Recently, a construction for an FL* was proposed starting from a set of decoupled wires in ref. \cite{Tsvelik16} . However, the phase obtained in the above paper is not a $\mathbb{Z}_2$ FL* and does not discuss the topological structure or nature of its anyonic excitations.}}.

\subsection{Kondo Hamiltonian}

We begin by using a local Kondo-coupling, $H_K = J_K~\sum_j \left[\mathbf{S}\cdot\mathbf{s}\right]_j$, that preserves the SU(2) spin rotation symmetry. Then
\beq
H_{FL^\ast} &=& H_{el} + H_{SL} + H_K,  \nonumber \\
H_K &=& \frac{J_K}{4}\sum_{j,\sigma\sigma^\prime}\int dx~\mathbf{\Gamma}_j \cdot \mathbf{\tau}_j, \nonumber \\
\mathbf{\Gamma}_j  &=& \left(\psi_{j\sigma}^{R\dagger}\boldsymbol{\tau}_{\sigma\sigma^\prime} \psi_{j\sigma^\prime}^R + \psi_{j\sigma}^{L\dagger}\boldsymbol{\tau}_{\sigma\sigma^\prime} \psi_{j\sigma^\prime}^L\right), \nonumber \\
\mathbf{\tau}_j &=& \left(c_{j\sigma}^{R\dagger}\boldsymbol{\tau}_{\sigma\sigma^\prime} c_{j\sigma^\prime}^R + c_{j\sigma}^{L\dagger}\boldsymbol{\tau}_{\sigma\sigma^\prime} c_{j\sigma^\prime}^L\right).
\eeq
where $H_{SL}$ is as described in Eqn.~\ref{hsl} earlier.  Even though the Hamiltonian looks like a standard Kondo-type Hamiltonian, there is a subtlety associated here with the specific construction used to arrive at the description of the $\mathbb{Z}_2$ spin-liquid. The  spin-spin coupling in $H_K$ does not commute with $H_{IC}$ (in $H_{SL}$; see Eqn. \ref{hsl}) as $S^{x,y}_j$ depend on $\hat{\phi}_j$ after bosonization. However, we appeal to our physical intuition here; since the spin liquid background is $gapped$, the phase obtained by coupling it to a fermi liquid will be perturbatively stable as long as the Kondo-coupling is small compared to the typical gaps (i.e. $J_K\ll \tn{min}\{\Delta_{1,2}\}$). Thus in the small $J_K$ limit we realize the $\mathbb{Z}_2$ FL* phase without any broken symmetries. However, it remains an interesting open problem to study the fate of this phase when the above condition is not satisfied. 

\subsection{Ising limit}
There is a special limit in which the complications described above can be circumvented. Suppose if $J_K^z\gg J_K^{x,y}$ as a result of easy-axis anisotropy. The Kondo-coupling then essentially involves only a local $H_K^z = J_K \sum_j\left[S^zs^z\right]_j$ coupling. We then have
\beq
H_{FL^\ast} &=& H_{el} + H_{SL} + H_K^z,  \nonumber \\
H_K^z &=&\frac{J_K^z}{4}\sum_j\int dx~\mathbf{\Gamma}_j^z~\mathbf{\tau}_j^z.
\eeq
Since $S^z_j$ depends only on $\hat{\theta}_j$ in the bosonized language, it commutes with $H_{IC}$ (in $H_{SL}$; see eqn. \ref{hsl}). Thus, the spin liquid background is stable for any reasonable value of $J_K$, as long as it is not strong enough to drive Kondo screening. Moreover, we do not expect the gapped spinons to induce any non-Fermi liquid behavior for the electrons.
Therefore, for small values of $J_K$ we realize once again a $\mathbb{Z}_2$ FL* (that explicitly breaks the SU(2) spin rotation symmetry) with a Fermi surface of the type shown in Fig.~\ref{cwire}(b). 

\section{Discussion}
\label{conc}
In this work, we have tried to extend the general program of constructing two dimensional correlated phases of matter by coupling together an array of one dimensional wires. In particular, we have demonstrated that it is possible to construct explicitly a time-reversal symmetric phase of matter that has the following characteristics: (i) Energy gap in the bulk and at the edge, (ii) three bulk anyonic quasiparticles which are mutual semions, (iii) non-trivial ground state degeneracy on a torus, and, is the $\mathbb{Z}_2$ spin-liquid. In the limit of a weak `Kondo'-type coupling to an itinerant Fermi-sea, we have also constructed a $\mathbb{Z}_2$ FL*.

It would be interesting to explore the possibility of realizing other time-reversal symmetric spin-liquid phases in two spatial dimensions with gapless excitations in the bulk. A particular example is the $U(1)$ spin-liquid with a spinon fermi-surface \cite{SSL08}, which can potentially be constructed in a manner similar to the one proposed for the half-filled Landau level \cite{DM16}. The fate, or even the existence, of the strong-coupling fixed point for the above spin-liquid problem remains unanswered \cite{SSLee} and it would be interesting to see if a complementary approach, such as the one proposed here, can address some of these unresolved questions.    

\begin{acknowledgements}
We thank S. Sachdev for helpful suggestions and discussions that inspired this work and M. Barkeshli for useful discussions. AAP is supported by the NSF under Grant DMR-1360789. DC is supported by a postdoctoral fellowship from the Gordon and Betty Moore Foundation, under the EPiQS initiative, Grant GBMF-4303 at MIT. We also acknowledge support by the ``2016 Boulder Summer School on Condensed Matter Physics --- Topological phases of quantum matter" through NSF grant DMR-13001648 during which part of this work was completed.
\end{acknowledgements}

\bibliographystyle{apsrev4-1_custom}
\bibliography{cwc}

\appendix

\end{document}